\documentclass[11pt,letterpaper]{article}
\providecommand{\hmajdir}{}
\IfFileExists{arxiv-hmaj/hmaj-preamble.tex}{\renewcommand{\hmajdir}{arxiv-hmaj/}}{}
\usepackage[T1]{fontenc}
\usepackage[utf8]{inputenc}
\usepackage[margin=1in,letterpaper]{geometry}
\usepackage{authblk}
\usepackage{amsmath,amssymb,amsthm}
\usepackage{lmodern}
\usepackage{mathtools}
\usepackage{dsfont}
\makeatletter
\@ifundefined{newcounteralias}{%
  \usepackage{thmtools}
  \declaretheorem[name=Theorem]{theorem}
  \declaretheorem[name=Lemma,sibling=theorem]{lemma}
  \declaretheorem[name=Corollary,sibling=theorem]{corollary}
  \declaretheorem[name=Remark,sibling=theorem,style=remark]{remark}
}{%
  \newtheorem{theorem}{Theorem}
  \newtheorem{lemma}[theorem]{Lemma}
  
  \theoremstyle{remark}
  
  \usepackage{thmtools}
}
\makeatother
\usepackage[dvipsnames]{xcolor}
\usepackage{xspace}
\usepackage{enumitem}
\usepackage{microtype}
\usepackage{url}
\usepackage[colorlinks=true,linkcolor=BrickRed,citecolor=BrickRed,urlcolor=BrickRed]{hyperref}
\usepackage[capitalize,nameinlink]{cleveref}
\allowdisplaybreaks

\newcommand{\dejavu}{\textnormal{\textsc{DéjàVu}}\xspace}
\newcommand{\dejavuh}{\textnormal{\textsc{DéjàVu}}(\(h\))\xspace}
\newcommand{\dejavuof}[1]{\textnormal{\textsc{DéjàVu}}(\(#1\))\xspace}
\newcommand{\hmaj}{\(h\)-\textnormal{\textsc{Majority}}\xspace}
\newcommand{\threemaj}{\(3\)-\textnormal{\textsc{Majority}}\xspace}
\newcommand{\twochoices}{\(2\)-\textnormal{\textsc{Choices}}\xspace}
\newcommand{\pullh}{\(\mathcal{PULL}(h)\)\xspace}
\newcommand{\pullof}[1]{\(\mathcal{PULL}(#1)\)\xspace}

\newcommand{\E}{\mathbb{E}}
\newcommand{\Z}{\mathbb{Z}}
\newcommand{\Var}{\operatorname{Var}}
\newcommand{\one}{\mathds{1}}
\newcommand{\abs}[1]{\lvert #1\rvert}
\newcommand{\whp}{w.h.p.\xspace}

\title{Plurality Consensus for \(h\)-Majority via First-Repetition Rules}
\author[1]{Niccolò D'Archivio}
\author[2]{George Giakkoupis}
\author[1]{Frédéric Giroire}
\author[1]{Emanuele Natale}
\affil[1]{COATI, Inria, CNRS, I3S, Université Côte d'Azur, France}
\affil[2]{Inria, Univ Rennes, CNRS, IRISA, France}
\date{}

\begin{document}
\maketitle

\begin{abstract}
In the \hmaj dynamics, each of \(n\) agents holds one of \(k\) opinions and, in every synchronous
round, observes the opinions of \(h\) agents chosen independently and uniformly at random and adopts a
most frequent one among them, breaking ties uniformly at random. Write \(M_0\) for the initial number
of agents holding the most frequent opinion, and \(\Delta_0\) for its bias, its lead over the second most
frequent opinion. We show that, for every sample size \(h\ge3\), if
\(\Delta_0\ge L\sqrt{M_0\log n}\cdot\max\{1,n/(h^2M_0)\}\) for a sufficiently large constant \(L\), then
\hmaj reaches consensus on the initially most frequent opinion within \(O((1+n/(h^2M_0))\log n)\) rounds
with high probability. This is the time bound proved for the \dejavu dynamics with \(h\) samples; our
bias condition is that of \dejavu when \(h\ge\sqrt{n/M_0}\), and larger by a factor
\(\sqrt{n/(h^2M_0)}\) otherwise. Our main contribution is the analysis of one round, which rests on a comparison principle:
among all rules that read the same observations and treat opinions symmetrically, plurality maximizes
the difference between the probability of outputting the most frequent opinion and that of outputting
any fixed competitor. We apply it to the rule that adopts the first opinion repeated among \(h-1\)
observations, and otherwise the \(h\)-th observation (which stands in for the agent's own opinion); its
output law is the law of the new opinion of a uniformly random agent after one round of \dejavu with
\(h-1\) samples. Hence, in expectation, one round of
\hmaj increases the gap between the most frequent opinion and each competitor at least as much as that
round of \dejavu, whose analysis we use as a black box. Independently and
concurrently, d'Amore, Rybicki, Shimizu, Shiraga and Vacus obtained, among other results, the same time
bound under the weaker bias condition of \dejavu, \(\Delta_0=\Omega(\sqrt{\max\{M_0,n/h^2\}\log n})\),
with a different
analysis of one round and a martingale analysis over many rounds.
\end{abstract}

\section{Introduction}\label{sec:intro}

Plurality consensus asks a population of \(n\) agents, initially split among \(k\) opinions, to converge
to the opinion with the largest initial support~\cite{becchettiConsensusDynamicsOverview2020}. In the \hmaj
dynamics~\cite{becchettiSimpleDynamicsPlurality2017}, in every synchronous round each agent observes
the opinions of \(h\) agents chosen independently and uniformly at random, and adopts a most frequent
opinion among its \(h\) observations, breaking ties uniformly at random. Observing \(h\) random agents
per round is the \pullh model~\cite{DoerrGMSS11,BecchettiCNPT16,becchettiSimpleDynamicsPlurality2017}.
For \(h=3\), \hmaj is \threemaj; for \(h\le2\), it has the transition law of the voter
dynamics~\cite{hassin2001distributed}, which does not favor any opinion.

\threemaj and the related \twochoices dynamics~\cite{cooper2014power,elsasser2016efficient}, in which an
agent adopts the opinion of two random agents if they agree and otherwise keeps its own, are well
understood. From an arbitrary
initial configuration, \threemaj reaches consensus within \(\tilde O(\min\{k,\sqrt n\})\) rounds and
\twochoices within \(\tilde O(k)\) rounds~\cite{Shimizu2025}, improving
on~\cite{becchettiSimpleDynamicsPlurality2017,BerenbrinkCEKMN17,GhaffariL18}; moreover, up to
logarithmic factors, the consensus time of either dynamics depends on the initial configuration only
through the largest initial opinion density~\cite{darchivio2026consensus}.
For general \(h\), Becchetti et al.~\cite{becchettiSimpleDynamicsPlurality2017} gave a lower bound of
\(\Omega(k/h^2)\) rounds for \hmaj, valid for suitable initial configurations close to the uniform one
whenever \(k/h=O(n^{1/4-\varepsilon})\) for some constant \(\varepsilon>0\).

Let \(M_0\) be the initial count of the most frequent opinion, and let \(\Delta_0\), its \emph{bias}, be
the difference between \(M_0\) and the initial count of the second most frequent opinion. Apart from the
concurrent work discussed below, the only upper bound for \hmaj that allows both \(h\) and \(k\) to grow
requires very large samples: if \(M_0=\omega(\log n)\), \(\Delta_0=\omega(\sqrt{M_0})\) and
\(h=\omega(n\log n/M_0)\), then \hmaj reaches consensus on the initially most frequent opinion within
\(O(\log n)\) rounds with high probability~\cite{damoreHMajorityDynamicsMany2025}. From almost balanced
configurations, in which \(M_0\) is of order \(n/k\), this condition on the sample size reads
\(h=\omega(k\log n)\).

The \dejavu dynamics~\cite{damore2026dejavuba,damore2026dejavuminimalisticmechanismdistributed}
replaces counting by repetition detection. In its variant \dejavuh in \pullh, an agent reads its \(h\)
observations of a round in order and adopts the first opinion that it observes twice; if the \(h\)
observations are pairwise distinct, it keeps its opinion. d'Amore et
al.~\cite[Theorem~1]{damore2026dejavuminimalisticmechanismdistributed} showed that, provided
\(M_0=\omega(\log n)\) and \(\Delta_0=\Omega(\sqrt{\max\{M_0,n/h^2\}\log n})\), \dejavuh brings all
agents to the initial plurality opinion within \(O((1+n/(h^2M_0))\log n)\) rounds with high
probability. Upper bounds of this
kind for \hmaj with smaller samples were left open there.

\paragraph{Our results.}
We show that, for every \(h\ge3\), \hmaj has the time bound of \dejavuh (\cref{thm:main}). If the bias
satisfies
\[
 \Delta_0\ge L_c\sqrt{M_0\log n}\cdot\max\Bigl\{1,\frac n{h^2M_0}\Bigr\}
\]
for a sufficiently large constant \(L_c\), then \hmaj reaches consensus on the initially most frequent
opinion within
\[
 O\Bigl(\Bigl(1+\frac n{h^2M_0}\Bigr)\log n\Bigr)
\]
rounds with high probability. For \(h\ge\sqrt{n/M_0}\), this bias condition reads
\(\Delta_0\ge L_c\sqrt{M_0\log n}\), as for \dejavuh; for smaller \(h\), it exceeds the condition for
\dejavuh by the factor \(\sqrt{n/(h^2M_0)}\).

The main contribution of this note is the analysis of one round: a comparison principle that reduces
one round of \hmaj to one round of \dejavuof{h-1}, whose
analysis~\cite[Lemma~23]{damore2026dejavuminimalisticmechanismdistributed} we use as a black box. It
differs from the analysis of one round in the concurrent work discussed below. We derive
\cref{thm:main} from the one-round estimates with a short round-by-round argument.

\paragraph{Concurrent work.}
Independently and concurrently, d'Amore, Rybicki, Shimizu, Shiraga and Vacus proved the time bound of
\cref{thm:main} under the weaker bias condition
\[
 \Delta_0=\Omega\Bigl(\sqrt{\max\Bigl\{M_0,\frac n{h^2}\Bigr\}\log n}\Bigr)
\]
\cite[Theorem~1.2]{damore2026nearlyoptimal}.\footnote{Their
detailed statement~\cite[Theorem~6.1]{damore2026nearlyoptimal} has success probability at least
\(1-1/n\); our proof gives \(1-n^{-c}\) for every fixed \(c>0\), with \(L_c\) depending on \(c\).} They
also proved that, for every \(h\ge3\) and from every initial configuration, \hmaj reaches consensus
within
\[
 \tilde O\Bigl(1+\min\Bigl\{\frac1{\min\{h^2M_0/n,1\}},\frac{\sqrt n}h\Bigr\}\Bigr)
\]
rounds~\cite[Theorem~1.1]{damore2026nearlyoptimal}. Their lower bound for neutral \(h\)-input dynamics
(fixed, possibly randomized rules that depend only on the \(h\) observations and the agent's current
opinion, output one of these \(h+1\) opinions, and treat opinions symmetrically) matches this bound, as a function of \(n\), \(h\) and \(M_0\)
and for \(M_0<n\), up to polylogarithmic factors~\cite[Theorem~1.3]{damore2026nearlyoptimal}.

The two analyses differ in the one-round estimate. Theirs bounds the growth of the relative gap
\(q_1/q_j-1\) between the output probabilities \(q_1\) of the most frequent opinion and \(q_j\) of a
competitor~\cite[Lemma~3.1]{damore2026nearlyoptimal}, through probability-transport arguments on the
vectors of sample counts. Ours bounds the growth of the difference \(q_1-q_j\), by comparing plurality
with a first-repetition rule, which reduces the estimate to the one-round analysis of
\dejavu~\cite{damore2026dejavuminimalisticmechanismdistributed}. Their multi-round analysis follows the
martingale framework based on Freedman's inequality that was developed for \threemaj by Cooper,
Mallmann-Trenn, Radzik, Shimizu and Shiraga~\cite{CooperMRSS25} (in an asynchronous setting) and for
\threemaj and \twochoices by Shimizu and Shiraga~\cite{Shimizu2025}.

Both works prove that plurality does not decrease the ratio of the most frequent opinion to any
other, \(q_j/q_1\le p_j/p_1\) (\cref{lem:boost}
and~\cite[Corollary~5.12]{damore2026nearlyoptimal}). They also compare one round of \hmaj with one
round of plurality on fewer observations, through a majorization of output
laws~\cite[Lemma~5.15 and Corollary~5.18]{damore2026nearlyoptimal}, which they use in their final phase with three
observations~\cite[Lemma~6.9]{damore2026nearlyoptimal}, as we do in \cref{lem:one-round}(e). That
comparison is between plurality rules with different numbers of observations. \Cref{lem:compare}
compares plurality with every label-equivariant rule on the same observations, which is what makes the
first-repetition rule, and hence \dejavu, available. Write \(q^{(m)}\) for the output law of plurality on \(m\) observations, with opinion \(1\) a
most frequent opinion. Applied to plurality on the first \(m<h\) observations, \cref{lem:compare} gives
\(q^{(h)}_1\ge q^{(m)}_1\), the first prefix inequality of their majorization, but not the others.

Our analysis of one round (the comparison principle of \cref{lem:compare} and its application to
first-repetition rules) was written up in a draft dated 26 September 2026, before the work of d'Amore
et al.~\cite{damore2026nearlyoptimal} appeared on arXiv on 29 October 2026.

\paragraph{Approach.}
The main step is a lower bound on the expected growth, in one round, of the gap between the count of
the most frequent opinion and that of each competitor. A heuristic gives the rate of this growth. If
opinion \(1\) has density \(p_1\) and a competitor \(j\) has density \(p_j\), a given pair of
observations shows opinion \(1\) twice with probability \(p_1^2\) and opinion \(j\) twice with
probability \(p_j^2\). Among \(h\) observations there are \(\binom h2\) pairs, so repetitions favor
opinion \(1\) over \(j\) by about
\[
 \binom h2\bigl(p_1^2-p_j^2\bigr)\approx h^2p_1\,(p_1-p_j),
\]
which is \(h^2p_1\) times the density gap, as long as \(h^2p_1\le1\). At the initial configuration,
\(p_1=M_0/n\), and this rate is the scale \(\min\{h^2M_0/n,1\}\) that appears in the bias condition and
in the time bound.

The output law of plurality on \(h\) observations is a sum over count vectors with no tractable closed form, so we do not
compute it.
Instead, we use the first-repetition rule \(D_h\), which reads the \(h\) observations in order and
outputs the first one among the first \(h-1\) that repeats an earlier one, or the last observation if
the first \(h-1\) are pairwise distinct. Its output law is the law of the next opinion of a uniformly
random agent in \dejavuof{h-1}. Indeed, an agent of \dejavuof{h-1} that sees no repetition among its
\(h-1\) observations keeps its current opinion, and the current opinion of a uniformly random agent is
distributed as one more observation, independent of the others. The analysis of
\dejavu~\cite[Lemma~23]{damore2026dejavuminimalisticmechanismdistributed} shows that \dejavuof{h-1}
multiplies, in expectation, the gap to every competitor by \(1+\Omega(\min\{h^2C_1/n,1\})\) as long
as \(C_1<3n/4\), where
\(C_1\) is the current count of the most frequent opinion. A comparison principle (\cref{lem:compare})
transfers this amplification to plurality (\cref{lem:dominate}): among all randomized rules that treat
opinions symmetrically and read the same observations, plurality maximizes the difference between the
probability of outputting the most likely opinion and that of outputting any fixed competitor.

Given these one-round estimates, the analysis over many rounds is short (\cref{sec:rounds}). Under
the bias condition, Bernstein's inequality shows that the fluctuations of one round are a small
fraction of the expected growth of every gap. Hence, with high probability, every round multiplies
the bias by \(1+\Omega(\min\{h^2M_0/n,1\})\), does not decrease the count of opinion \(1\), and
preserves the bias condition (\cref{lem:round}). Once at least \(3n/4\) agents hold opinion \(1\),
the expected number of the other agents decreases by a constant factor in every round (\cref{lem:endgame}).

\Cref{sec:model} defines the model and states the result, \cref{sec:one-round} analyzes one round of
\hmaj, and \cref{sec:rounds} proves \cref{thm:main}.

\section{Model, results and tools}\label{sec:model}

\paragraph{Configurations.}
There are \(n\) agents, each holding an opinion in the set \([k]=\{1,\dots,k\}\), where \(2\le k\le n\);
some opinions may be held by no agent, initially or later. The \emph{configuration} is \(C=(C_1,\dots,C_k)\), where \(C_i\) is
the number of agents holding opinion \(i\), and \(p=C/n\) is the vector of opinion densities. Primes
denote the next round (\(C'\) and \(p'\)). Opinion \(1\) is \emph{a largest opinion} if
\(C_1=\max_iC_i\). For every opinion \(j\ne1\) we write
\[
 \Delta_j=C_1-C_j,\qquad \Delta=\min_{j\ne1}\Delta_j=C_1-\max_{j\ne1}C_j .
\]
When \(\Delta>0\), opinion \(1\) is the unique largest opinion, \(\Delta\) is its \emph{bias}, and
\(\Delta/C_1\) is its \emph{relative bias}. Logarithms are natural. A sequence of events holds
\emph{with high probability} (\whp) if its probability is at least \(1-n^{-c}\), where \(c>0\) is fixed
but arbitrary; all constants implicit in \(O(\cdot)\) or in ``sufficiently large'' may depend on \(c\)
only, and not on \(n\), \(k\), \(h\) or the initial configuration.

\paragraph{\texorpdfstring{\hmaj in \pullh}{h-Majority in PULL(h)}.}
In the \pullh model, time is divided into synchronous rounds. In each round, every agent samples \(h\)
agents independently and uniformly at random (with replacement, so an agent, possibly itself, may be
sampled several times) and learns their current opinions; afterwards all agents update at the same
time. In \hmaj, an agent adopts a
most frequent opinion among its \(h\) observations, breaking ties uniformly at random and independently
of everything else; it ignores its current opinion. Since every adopted opinion has just been
observed, an opinion held by no agent never returns; in particular, consensus is absorbing.

Given a probability vector \(p\) on \([k]\) and an integer \(m\ge1\), draw \(m\) labels independently from
\(p\), and select uniformly at random one of the labels that occur most often; the \emph{output law of
plurality on \(m\) draws} is the law \(q=q(p)\) of the selected label. Given
the configuration \(C\), the \(h\) observations of an agent are independent draws from \(p=C/n\),
independently across agents. Hence, given \(C\), the next opinions of the agents of \hmaj are
independent, each with law \(q(p)\) for \(m=h\). In particular, given \(C\), the next configuration
\(C'\) has the multinomial distribution with parameters \(n\) and \(q(p)\); this is the
\emph{transition law} of \hmaj in \pullh. Its mean is
\begin{equation}\label{eq:output}
 \E[C_i'\mid C]=n\,q_i(p)\qquad(i\in[k]).
\end{equation}

\subsection{Result}\label{sec:results}

\begin{theorem}[\hmaj in \pullh]\label{thm:main}
For every constant \(c>0\) there are constants \(L_c,A_c>0\), depending only on \(c\), such that the
following holds. Let \(n\ge2\) and \(2\le k\le n\), and let \(h\ge3\) be an integer, which may depend on
\(n\) and on the initial configuration. Suppose that opinion \(1\) is initially the unique largest
opinion, with count \(M_0\) and bias \(\Delta_0\). If the \emph{bias condition}
\begin{equation}\label{eq:bias-condition}
 \Delta_0\ge L_c\sqrt{M_0\log n}\cdot\max\Bigl\{1,\frac n{h^2M_0}\Bigr\}
\end{equation}
holds, then, with probability at least \(1-n^{-c}\), \hmaj in \pullh reaches consensus on opinion \(1\)
within
\[
 A_c\Bigl(1+\frac n{h^2M_0}\Bigr)\log n
\]
rounds.
\end{theorem}

Both quantities are governed by the \emph{amplification scale}
\[
 \alpha(x)=\min\Bigl\{\frac{h^2x}n,1\Bigr\}\qquad(x\ge0).
\]
Since \(1/\alpha(M_0)=\max\{1,n/(h^2M_0)\}\), the bias condition reads
\[
 \alpha(M_0)\,\Delta_0\ge L_c\sqrt{M_0\log n},
\]
and the time bound is \(O(\log n/\alpha(M_0))\). For \(h\ge\sqrt{n/M_0}\), we have \(\alpha(M_0)=1\): the
bias condition reads \(\Delta_0\ge L_c\sqrt{M_0\log n}\), and the time bound is \(O(\log n)\). Since
\(\Delta_0\le M_0\) and \(\alpha(M_0)\le1\), the bias condition implies \(M_0\ge L_c^2\log n\), so the
theorem needs no separate lower bound on \(M_0\). In our proof, \(h\ge3\) is used twice: \hmaj is compared with \dejavu
with \(h-1\ge2\) samples (\cref{lem:dominate}), and, once \(C_1\ge3n/4\), with plurality on three of its
observations (\cref{lem:one-round}(e)). The restriction \(h\ge3\) is needed: for
\(h\in\{1,2\}\), \(q(p)=p\) (for \(h=2\), \(p_i^2+p_i(1-p_i)=p_i\)), so the count of opinion \(1\) is a
bounded martingale, and consensus, which is reached almost surely, is on opinion \(1\) with probability
\(M_0/n\).

\subsection{Bernstein's inequality}\label{sec:tools}

For a sum \(X\) of independent random variables \(Y_1,\dots,Y_N\) with \(\abs{Y_i-\E Y_i}\le J\) and
\(\Var X\le V\), Bernstein's inequality~\cite[Corollary~2.11]{boucheron2013concentration}, applied to
\(X\) and to \(-X\), states that \(\Pr(\abs{X-\E X}\ge z)\le2\exp(-z^2/(2(V+Jz/3)))\) for every
\(z>0\). We use it in the following form.

\begin{lemma}[Bernstein's inequality]\label{lem:bernstein}
Let \(X=Y_1+\dots+Y_N\) be a sum of independent random variables with \(\abs{Y_i-\E Y_i}\le J\) and
\(\Var X\le V\). Then, for every \(y>0\),
\[
 \Pr\bigl(\abs{X-\E X}\ge\sqrt{2Vy}+Jy\bigr)\le2e^{-y}.
\]
\end{lemma}

\begin{proof}
Let \(z=\sqrt{2Vy}+Jy\). Then
\[
 z^2-2Vy=2Jy\sqrt{2Vy}+J^2y^2\ge\frac23Jy\bigl(\sqrt{2Vy}+Jy\bigr)=\frac23Jyz,
\]
that is, \(z^2/(2(V+Jz/3))\ge y\), and Bernstein's inequality gives the claim.
\end{proof}

\section{One round of \texorpdfstring{\hmaj}{h-Majority}}\label{sec:one-round}

Throughout this section, \(p\) is a probability vector on \([k]\), and \(q=q(p)\) is the output law of
plurality on \(h\) independent draws from \(p\); by~\eqref{eq:output}, \(nq\) is the expected
configuration after one round of \hmaj from \(C=np\). The goal of the section is \cref{lem:one-round},
which collects the one-round properties of \hmaj used in \cref{sec:rounds}.

\subsection{Majority boosting}

The first property is that plurality preserves the order of the densities and can only increase their
ratios. In the terminology of d'Amore et
al.~\cite{damore2026dejavuminimalisticmechanismdistributed}, who show that the squared Euclidean norm of the
configuration is a submartingale for every such dynamics~\cite[Lemma~35]{damore2026dejavuminimalisticmechanismdistributed},
\hmaj with uniform tie-breaking is \emph{majority boosting}. The next lemma and its proof also appear
in~\cite{darchivio2026dejavufirst}, and its claims for a largest label \(1\) are also
proved, for \(h\ge3\), in the concurrent work~\cite[Corollary~5.12]{damore2026nearlyoptimal}; we include the proof to keep the note
self-contained. For a vector \(x\in\Z_{\ge0}^k\) of counts
with \(\sum_\ell x_\ell=h\), the labels attaining \(\max_\ell x_\ell\) form a set \(\mathcal W(x)\), and
label \(i\) wins the uniform tie-break with probability
\(w_i(x)=\one[i\in\mathcal W(x)]/\abs{\mathcal W(x)}\). Writing
\(P(x)=\frac{h!}{\prod_\ell x_\ell!}\prod_\ell p_\ell^{x_\ell}\) (with \(0^0=1\)) for the multinomial
probability of \(x\) under \(h\) draws from \(p\), we have \(q_i=\sum_xP(x)w_i(x)\).

\begin{lemma}[Majority boosting]\label{lem:boost}
Let \(h\ge1\), and let \(i,j\) be labels with \(p_i\ge p_j\). Then \(p_jq_i\ge p_iq_j\) and \(q_i\ge q_j\).
If moreover \(p_1=\max_\ell p_\ell\), then
\[
 q_1\ge p_1\qquad\text{and}\qquad q_1-q_j\ge\frac{q_1}{p_1}(p_1-p_j)\quad\text{for every label }j.
\]
\end{lemma}

\begin{proof}
We first prove \(p_jq_i\ge p_iq_j\). This is immediate if \(i=j\), or if \(p_j=0\), since then label
\(j\) is never drawn and \(q_j=0\). Otherwise write \(u=p_i\ge v=p_j>0\) and
\[
 d(x)=P(x)\bigl(vw_i(x)-uw_j(x)\bigr),\qquad\text{so that}\qquad vq_i-uq_j=\sum_xd(x).
\]
We group the count vectors so that each group has a nonnegative total contribution. For a count vector
\(x\), let \(\tilde x\) be \(x\) with its \(i\)- and \(j\)-counts exchanged. With the counts
\(x_\ell\), \(\ell\notin\{i,j\}\), fixed, we have \(P(x)=F(x_i,x_j)\), where
\[
 F(a,b)=\frac{\kappa u^av^b}{a!\,b!},\qquad\kappa=h!\prod_{\ell\notin\{i,j\}}\frac{p_\ell^{x_\ell}}{x_\ell!}\ge0 .
\]

\emph{Unequal counts.} Pair each count vector \(x\) with \(x_i\ne x_j\) with its image \(\tilde x\), choosing the
representative with \(x_i=a>b=x_j\). Label \(j\) is not of maximal count in \(x\), and exchanging the
two counts exchanges the selection probabilities of \(i\) and \(j\): \(w_j(x)=0\), \(w_i(\tilde x)=0\)
and \(w_j(\tilde x)=w_i(x)\). Thus
\[
 d(x)+d(\tilde x)=w_i(x)\bigl(vF(a,b)-uF(b,a)\bigr)
 =\frac{\kappa w_i(x)}{a!\,b!}(uv)^{b+1}\bigl(u^{a-b-1}-v^{a-b-1}\bigr)\ge0,
\]
since \(u\ge v\) and \(a-b-1\ge0\).

\emph{Equal counts.} Let \(x_i=x_j=a\). If \(i\) and \(j\) are not of maximal count, then \(d(x)=0\).
Otherwise both belong to \(\mathcal W(x)\), so \(\abs{\mathcal W(x)}\ge2\) and
\(w_i(x)=w_j(x)=1/\abs{\mathcal W(x)}\); moreover \(a\ge1\), since a label of maximal count among
\(h\ge1\) draws is drawn at least once. Then
\[
 d(x)=F(a,a)\,\frac{v-u}{\abs{\mathcal W(x)}}=-\frac{\kappa(uv)^a(u-v)}{(a!)^2\,\abs{\mathcal W(x)}}\le0 .
\]
To offset this term, let \(y\) be \(x\) with one draw moved from \(j\) to \(i\), so \(y_i=a+1\) and
\(y_j=a-1\). Label \(i\) is the unique label of maximal count in \(y\), so \(w_i(y)=1\), and the
computation for unequal counts, with \(a+1\) and \(a-1\) in place of \(a\) and \(b\), gives
\[
 d(y)+d(\tilde y)=\frac{\kappa(uv)^a(u-v)}{(a+1)!\,(a-1)!}
 =\frac{a\abs{\mathcal W(x)}}{a+1}\bigl(-d(x)\bigr)\ge-d(x),
\]
since \(a\abs{\mathcal W(x)}\ge2a\ge a+1\). Distinct such \(x\) give distinct pairs \(\{y,\tilde y\}\), because \(x\)
is recovered from \(\{y,\tilde y\}\) by replacing the \(i\)- and \(j\)-counts by their average.

Every count vector now lies in exactly one of the following groups: a vector \(x\) with \(x_i=x_j\) in
which \(i\) and \(j\) are of maximal count, together with its pair \(\{y,\tilde y\}\); a remaining pair
\(\{x,\tilde x\}\) with \(x_i\ne x_j\); a vector with \(x_i=x_j\) in which \(i\) and \(j\) are not of
maximal count. Every group has a nonnegative total contribution, hence \(vq_i-uq_j\ge0\).

If \(p_j>0\), this gives \(q_i\ge(p_i/p_j)q_j\ge q_j\); if \(p_j=0\), then \(q_j=0\le q_i\). Now let
\(p_1=\max_\ell p_\ell\), so that \(p_1>0\). Summing \(p_jq_1-p_1q_j\ge0\) over all labels \(j\) gives
\(q_1-p_1\ge0\). Finally, \(p_1q_j\le p_jq_1\) gives \(q_1-q_j\ge q_1-(p_j/p_1)q_1=(q_1/p_1)(p_1-p_j)\).
\end{proof}

\subsection{Plurality maximizes the gap to every competitor}

A rule that reads \(m\) labels and outputs a label, possibly using independent randomness, is
\emph{label-equivariant} if renaming the labels of its input renames its output in the same way: for
every permutation \(\psi\) of \([k]\) and every \(s\in[k]^m\), its output on
\(\psi(s)=(\psi(s_1),\dots,\psi(s_m))\) has the same law as \(\psi\) applied to its output on \(s\). Plurality with uniform tie-breaking is label-equivariant, since relabeling permutes the set of
labels of maximal count. The \emph{output law} of a rule on \(p\) is the law of its output when it reads
\(m\) independent draws from \(p\).

\begin{lemma}[Plurality maximizes the gap to every competitor]\label{lem:compare}
Let \(p_1=\max_\ell p_\ell\) and \(m\ge1\). Let \(r\) be the output law on \(p\) of a label-equivariant
rule that reads \(m\) labels, and let \(q\) be the output law of plurality on \(m\) draws from \(p\).
Then, for every label \(j\),
\[
 r_1-r_j\le q_1-q_j ;
\]
consequently, \(r_1\le q_1\).
\end{lemma}

Plurality on \(m\) multinomial draws is the classical single-sample procedure for selecting the most
probable cell~\cite{bechhofer1959single}. The consequence \(r_1\le q_1\) is the analogue, for one multinomial sample, of the
classical optimality of the natural selection rule among rules that treat the candidates symmetrically, which goes
back to Bahadur and Goodman~\cite{bahadur1952impartial} for independent samples from \(k\) populations;
see also~\cite{lehmann1966theorem}. The inequality \(r_1-r_j\le q_1-q_j\) is what the upper bound needs:
to show that plurality amplifies the gap between opinion \(1\) and a competitor \(j\), it suffices to
exhibit one label-equivariant rule on the same draws that does.

The proof writes \(r_1-r_j\) as the expected reward of a rule that earns \(+1\) for outputting opinion
\(1\), \(-1\) for outputting \(j\), and \(0\) otherwise. This reward is not monotone in the
probabilities: if \(p_t<p_j\) for some \(t\notin\{1,j\}\), then \(t\) earns more than the more likely
opinion \(j\). Therefore it does not suffice to compare two labels by exchanging the probabilities they
carry, and case~(c) of the proof groups the assignments of probabilities to three labels; it uses that the label output by plurality has the largest
count among all observed labels, not only a count at least that of the label it is compared with.

\begin{proof}
The second claim follows from the first: summing \(r_1-r_j\le q_1-q_j\) over all labels \(j\in[k]\) and
using \(\sum_jr_j=\sum_jq_j=1\) gives \(kr_1-1\le kq_1-1\). For the first claim, the case \(j=1\) is
trivial; let \(j\ne1\). We interpret both sides as the expected reward in a guessing game, and show that
plurality plays this game optimally after every observation.

\emph{A guessing game.} Call the indices \(1,\dots,k\) \emph{roles}: role \(t\) carries probability
\(p_t\). Choose a uniformly random permutation \(\pi\) of \([k]\), with the convention that label
\(\ell\) carries role \(\pi(\ell)\), and, given \(\pi\), draw \(S=(S_1,\dots,S_m)\) independently with
\(\Pr(S_b=\ell\mid\pi)=p_{\pi(\ell)}\). Given \(\pi\), the relabeled sequence \(\pi(S)\) consists of \(m\)
independent draws from \(p\). Let \(Q\) be the output of the rule on \(S\). Since the rule is
label-equivariant, given \(\pi\) the label \(\pi(Q)\) has the law of the output of the rule on
\(\pi(S)\), so \(\Pr(\pi(Q)=t\mid\pi)=r_t\) for every role \(t\). Naming label \(\ell\) earns the reward
\(\eta_{\pi(\ell)}\), where \(\eta_1=1\), \(\eta_j=-1\) and \(\eta_t=0\) for \(t\notin\{1,j\}\). The
expected reward of the rule is therefore \(\E[\eta_{\pi(Q)}]=r_1-r_j\), and the same argument for
plurality gives \(q_1-q_j\).

\emph{Posterior rewards.} Fix an observed sequence \(s\in[k]^m\) with count vector \(x\). Under the
assignment \(\pi\), its probability is \(L_\pi(x)=\prod_\ell p_{\pi(\ell)}^{x_\ell}\). Every \(\pi\) has
prior probability \(1/k!\), so Bayes' formula gives \(\Pr(\pi\mid S=s)=L_\pi(x)/L(x)\), where
\(L(x)=\sum_\tau L_\tau(x)\) is the total likelihood; sequences with \(L(x)=0\) have probability zero and
are ignored. The randomness of the rule is independent of \(\pi\) given \(S\). If the rule outputs label
\(i\) with probability \(\varrho_i(s)\) on seeing \(s\), its conditional expected reward given \(S=s\) is
therefore
\[
 \sum_i\varrho_i(s)\frac{\Psi_i(x)}{L(x)}\le\max_i\frac{\Psi_i(x)}{L(x)},\qquad\text{where}\quad
 \Psi_i(x)=\sum_\pi\eta_{\pi(i)}L_\pi(x).
\]
We show below that every label \(i\) of maximal count in \(x\) maximizes \(\Psi_i(x)\). Then all labels of
maximal count have the same, maximal, value of \(\Psi\), plurality attains \(\max_i\Psi_i(x)/L(x)\) after
every observation, and averaging over \(S\) gives \(r_1-r_j\le q_1-q_j\).

\emph{Grouping the assignments.} Fix a label \(i\) of maximal count in \(x\) and another label \(i'\).
Write \(a=x_i\ge b=x_{i'}\), \(u=p_1\) and \(v=p_j\); we use only that \(u\ge p_t\) for every role \(t\) (in
particular \(u\ge v\)), and no relation between \(v\) and the other probabilities. We show that
\[
 \Psi_i(x)-\Psi_{i'}(x)=\sum_\pi\bigl(\eta_{\pi(i)}-\eta_{\pi(i')}\bigr)L_\pi(x)\ge0
\]
by grouping the assignments so that each group has a nonnegative total contribution. Within each group
below, the roles of the labels outside a set \(B\) of at most three labels are fixed, and \(\kappa\ge0\)
denotes the product of the factors \(p_{\pi(\ell')}^{x_{\ell'}}\) over the labels \(\ell'\notin B\).
\begin{enumerate}[label=(\alph*),leftmargin=*]
\item If neither role \(1\) nor role \(j\) is carried by \(i\) or \(i'\), the contribution is zero.
\item If roles \(1\) and \(j\) are carried by \(i\) and \(i'\), group the two assignments that differ by
exchanging them (\(B=\{i,i'\}\)). Their contribution is
\[
 2\kappa\bigl(u^av^b-v^au^b\bigr)=2\kappa(uv)^b\bigl(u^{a-b}-v^{a-b}\bigr)\ge0 .
\]
\item Otherwise, exactly one of the roles \(1\) and \(j\) is carried by \(i\) or \(i'\). The other one is
carried by a third label \(\ell\), with count \(d=x_\ell\le a\), and the remaining label of
\(\{i,i'\}\) carries a role \(t\notin\{1,j\}\), with \(w=p_t\le u\). Fixing \(\ell\), \(t\) and the
roles outside \(B=\{i,i',\ell\}\) leaves four assignments:
\[
\begin{array}{c|c|c}
 (\pi(i),\pi(i'),\pi(\ell)) & \eta_{\pi(i)}-\eta_{\pi(i')} & \text{likelihood factor of }i,i',\ell\\ \hline
 (1,t,j) & +1 & u^aw^bv^d\\
 (t,1,j) & -1 & w^au^bv^d\\
 (j,t,1) & -1 & v^aw^bu^d\\
 (t,j,1) & +1 & w^av^bu^d
\end{array}
\]
Their total contribution is \(\kappa\Xi\), where
\[
 \Xi=w^b\bigl(u^av^d-v^au^d\bigr)-w^a\bigl(u^bv^d-v^bu^d\bigr).
\]
If \(b<d\),
the first bracket is \((uv)^d(u^{a-d}-v^{a-d})\ge0\) and the second is \((uv)^b(v^{d-b}-u^{d-b})\le0\),
so \(\Xi\ge0\). If \(b\ge d\), put \(\delta_1=a-b\) and \(\delta_2=b-d\); using \(w^a=w^bw^{\delta_1}\) and
\(u^{\delta_1+\delta_2}-v^{\delta_1+\delta_2}=u^{\delta_1}(u^{\delta_2}-v^{\delta_2})+v^{\delta_2}(u^{\delta_1}-v^{\delta_1})\),
\[
 \Xi=(uv)^dw^b\bigl[(u^{\delta_1}-w^{\delta_1})(u^{\delta_2}-v^{\delta_2})+v^{\delta_2}(u^{\delta_1}-v^{\delta_1})\bigr]\ge0 .
\]
\end{enumerate}
Each assignment belongs to exactly one group, so \(\Psi_i(x)\ge \Psi_{i'}(x)\). If \(i'\) also has maximal
count, exchanging the roles of \(i\) and \(i'\) gives \(\Psi_i(x)=\Psi_{i'}(x)\). No step divides by \(u\),
\(v\) or \(w\), so labels of probability zero are covered.
\end{proof}

\subsection{Comparison with \texorpdfstring{\dejavu}{DejaVu}}\label{sec:dejavu}

To apply \cref{lem:compare} we need a label-equivariant rule whose output law amplifies the gap. We
obtain one from the \dejavu dynamics~\cite{damore2026dejavuminimalisticmechanismdistributed}. For an
integer \(m\ge2\), in \dejavuof{m} in \pullof{m} every agent observes, in every synchronous
round, the opinions of \(m\) agents chosen independently and uniformly at random (with replacement, so an
agent may observe itself, as in \pullh); it reads these
observations in order and adopts the first opinion that repeats an earlier observation, and if the
\(m\) observations are pairwise distinct, it keeps its current
opinion~\cite[Section~3]{damore2026dejavuminimalisticmechanismdistributed}. The following one-round
estimate for this dynamics is~\cite[Lemma~23]{damore2026dejavuminimalisticmechanismdistributed}, which we
use as a black box.\footnote{In the notation of \cref{lem:dejavu} below:
\cite[Lemma~23]{damore2026dejavuminimalisticmechanismdistributed} states explicit constants; we only use that some absolute constant works. When \(m>k+1\), a repetition
occurs within the first \(k+1\) observations, so \dejavuof{m} coincides with \dejavuof{k+1}, and both
scales \(\min\{m^2C_1/n,1\}\) and \(\min\{(k+1)^2C_1/n,1\}\) equal \(1\), since \(kC_1\ge n\).}

\begin{lemma}[One round of \dejavu~{\cite[Lemma~23]{damore2026dejavuminimalisticmechanismdistributed}}]\label{lem:dejavu}
There is an absolute constant \(\gamma_1>0\) such that the following holds for every integer \(m\ge2\).
Let \(C\) be a configuration with \(C_1\ge C_2\ge\dots\ge C_k>0\) and \(C_1<3n/4\), and let
\(\widetilde C\) be the configuration after one round of \dejavuof{m} in \pullof{m} from
\(C\). Then, with \(\beta=\gamma_1\min\{m^2C_1/n,1\}\),
\[
 \E\bigl[\widetilde C_1-\widetilde C_j\bigm| C\bigr]\ge(1+\beta)\,(C_1-C_j)\quad(2\le j\le k),\qquad
 \E\bigl[\widetilde C_1\bigm| C\bigr]\ge C_1+\beta\,(C_1-C_2).
\]
\end{lemma}

The connection with \hmaj is a rule that reads \(h\ge3\) labels: the rule \(D_h\) reads
\(X_1,\dots,X_h\) in order and outputs the first label \(X_t\) with \(t\le h-1\) that repeats an earlier
one; if \(X_1,\dots,X_{h-1}\) are pairwise distinct, it outputs \(X_h\). It is label-equivariant:
relabeling the input by a permutation \(\psi\) maps the first label that repeats an earlier one to its
image under \(\psi\), and \(X_h\) to \(\psi(X_h)\). The rule \(D_h\) has the same output law as the rule that
outputs the first label that repeats an earlier one among all \(h\) draws, and a uniformly random draw
if the \(h\) draws are pairwise distinct. Indeed, both rules output the same label unless the \(h\)
draws are pairwise distinct; and, by exchangeability of the draws, the probability that the \(h\) draws
are pairwise distinct and the uniformly chosen draw is \(i\) equals the probability that they are pairwise
distinct and \(X_h=i\). So the \(h\)-th observation plays no special role in \(D_h\).

\begin{lemma}[\hmaj dominates \dejavuof{h-1}]\label{lem:dominate}
Let \(h\ge3\), and let opinion \(1\) be a largest opinion of a configuration \(C\). Let \(C'\) be the
configuration after one round of \hmaj in \pullh from \(C\), and let \(\widetilde C\) be the
configuration after one round of \dejavuof{h-1} in \pullof{h-1} from \(C\). Then, for every
label \(j\),
\[
 \E[C_1'-C_j'\mid C]\ge\E\bigl[\widetilde C_1-\widetilde C_j\bigm| C\bigr],\qquad
 \E[C_1'\mid C]\ge\E\bigl[\widetilde C_1\bigm| C\bigr].
\]
\end{lemma}

\begin{proof}
Let \(p=C/n\), let \(d\) be the output law of \(D_h\) on \(p\), and let \(q\) be the output law of
plurality on \(h\) draws from \(p\). For a label \(i\), let \(F_i\) be the probability that among
\(h-1\) independent draws from \(p\) some draw repeats an earlier one and the first such draw is \(i\),
and let \(N\) be the probability that the \(h-1\) draws are pairwise distinct. In \dejavuof{h-1}, the
observations of the agents are independent draws from \(p\), independent of the agents' current
opinions, and an agent adopts \(i\) if the first
repetition among its observations is \(i\), or if its observations are pairwise distinct and its
current opinion is \(i\). Summing over the agents, \(\E[\widetilde C_i\mid C]=nF_i+C_iN=n(F_i+p_iN)\).
The rule \(D_h\) outputs \(i\) if the first repetition among \(X_1,\dots,X_{h-1}\) is \(i\), or if
\(X_1,\dots,X_{h-1}\) are pairwise distinct and \(X_h=i\); as \(X_h\) is independent of
\(X_1,\dots,X_{h-1}\), \(d_i=F_i+Np_i\). Hence \(\E[\widetilde C_i\mid C]=nd_i\) for every label \(i\),
while \(\E[C_i'\mid C]=nq_i\) by~\eqref{eq:output}. Since \(D_h\) is label-equivariant and reads \(h\)
labels, \cref{lem:compare} gives \(d_1-d_j\le q_1-q_j\) and \(d_1\le q_1\), and multiplying by \(n\) gives
the claims.
\end{proof}

For \(h=3\), \dejavuof{2} is \twochoices, and \cref{lem:dominate} holds with equality: \(D_3\) and
plurality on three draws both have output law \(p_i(1+p_i-\sum_\ell p_\ell^2)\), so one round of
\threemaj and one round of \twochoices have the same expected effect. In general, the agent's current opinion, which
\dejavuof{h-1} keeps when it sees no repetition, has the same effect on the expected configuration as
a fresh observation, and \(D_h\) uses its \(h\)-th observation in this role.

\subsection{One-round estimates}

We now collect the properties of one round of \hmaj used in \cref{sec:rounds}. Let \(C\) be a
configuration in which opinion \(1\) is a largest opinion, so that \(C_1>0\), and let \(C'\) be the
configuration after one round of \hmaj in \pullh from \(C\). The expected growth factor of the count of
opinion \(1\) is
\[
 \rho=\rho(C)=\frac{\E[C_1'\mid C]}{C_1}=\frac{q_1}{p_1},
\]
and \(\alpha\) is the amplification scale of \cref{sec:results}. Finally, \(U=n-C_1\) and \(U'=n-C_1'\)
are the numbers of agents that do not hold opinion \(1\) in \(C\) and in \(C'\).

\begin{lemma}[One-round estimates]\label{lem:one-round}
There is an absolute constant \(\gamma_0\in(0,1]\) such that the following holds. Let \(h\ge3\), and let
opinion \(1\) be a largest opinion of \(C\). Then, for every \(j\ne1\):
\begin{enumerate}[label=(\alph*),leftmargin=*]
\item \emph{Ratio drift and order.} \(\rho\ge1\) and \(\E[\Delta_j'\mid C]\ge\rho\Delta_j\). Moreover,
\(C_i\ge C_{i'}\) implies \(\E[C_i'\mid C]\ge\E[C_{i'}'\mid C]\), for all labels \(i,i'\).
\item \emph{Variance.} Given \(C\), \(C_1'\) and \(\Delta_j'\) are sums of \(n\) independent terms, one per
agent, with values in \(\{0,1\}\) and in \(\{-1,0,1\}\) respectively, and
\[
 \Var(C_1'\mid C)\le\rho C_1,\qquad\Var(\Delta_j'\mid C)\le2\rho C_1 .
\]
\item \emph{Growth of opinion \(1\).} If \(C_1<3n/4\), then
\[
 \rho-1\ge\gamma_0\,\alpha(C_1)\,\frac\Delta{C_1} .
\]
\item \emph{Growth of the gaps.} If \(C_1<3n/4\), then
\[
 \E[\Delta_j'\mid C]\ge\max\bigl\{\rho,1+\gamma_0\,\alpha(C_1)\bigr\}\,\Delta ,
\]
and, if moreover \(C_j>0\), the same bound holds with \(\Delta_j\) in place of \(\Delta\).
\item \emph{Endgame.} If \(C_1\ge3n/4\), then
\[
 \E[U'\mid C]\le\frac{3U^2}n .
\]
\end{enumerate}
\end{lemma}

Part (c) says that opinion \(1\) grows in proportion to its relative bias \(\Delta/C_1\), at the rate
\(\alpha(C_1)\).

\begin{proof}
(a) By~\eqref{eq:output}, \(\E[C_i'\mid C]=nq_i\). Since \(p_1=\max_\ell p_\ell\), \cref{lem:boost} gives
\(q_1\ge p_1\), that is \(\rho\ge1\), and \(n(q_1-q_j)\ge(q_1/p_1)\,n(p_1-p_j)=\rho\Delta_j\). If
\(C_i\ge C_{i'}\), then \(p_i\ge p_{i'}\), and \cref{lem:boost} gives \(q_i\ge q_{i'}\).

(b) Given \(C\), the next opinions of the agents are independent with law \(q\) (\cref{sec:model}). The
term of an agent in \(C_1'\) is the indicator that its next opinion is \(1\), and its term in
\(\Delta_j'\) is that indicator minus the indicator that its next opinion is \(j\). Hence
\(\Var(C_1'\mid C)=nq_1(1-q_1)\le nq_1=\rho C_1\), and
\[
 \Var(\Delta_j'\mid C)=n\bigl[q_1+q_j-(q_1-q_j)^2\bigr]\le n(q_1+q_j)\le2nq_1=2\rho C_1,
\]
using \(q_j\le q_1\) from (a).

(c) and (d) Let \(\gamma_1\) be the constant of \cref{lem:dejavu}, and let
\(\gamma_0=\min\{4\gamma_1/9,1\}\). Let \(C_1<3n/4\), and let \(S\) be the set of opinions held by at least
one agent in \(C\). Opinions outside \(S\) are never observed, and every current opinion lies in \(S\), so they do not affect one round of
\dejavuof{h-1} from \(C\), which is one round of the same dynamics on the opinions in \(S\). Since
\(C_1<3n/4\), \(S\) contains an opinion other than \(1\), and \(\Delta=C_1-\max_{\ell\in S\setminus\{1\}}C_\ell\).
Renaming opinions does not change the transition law of \dejavu. Renaming the opinions in \(S\) as
\(1,\dots,\abs S\) in nonincreasing order of count, with opinion \(1\) first (possible since
\(C_1=\max_\ell C_\ell\)), we may therefore apply \cref{lem:dejavu} with \(m=h-1\ge2\) and \(\abs S\) in
place of \(k\), and there \(C_1-C_2=\Delta\); as \((h-1)^2\ge4h^2/9\) for \(h\ge3\), its parameter satisfies
\(\beta=\gamma_1\min\{(h-1)^2C_1/n,1\}\ge\gamma_0\,\alpha(C_1)\). Let \(\widetilde C\) be the configuration
after one round of \dejavuof{h-1} from \(C\).

For (c), \cref{lem:dominate,lem:dejavu} give
\(\rho C_1=\E[C_1'\mid C]\ge\E[\widetilde C_1\mid C]\ge C_1+\gamma_0\,\alpha(C_1)\,\Delta\).

For (d), the bound \(\E[\Delta_j'\mid C]\ge\rho\Delta_j\ge\rho\Delta\) is (a). If \(C_j>0\), then
\(j\in S\), and \cref{lem:dominate,lem:dejavu} give
\[
 \E[\Delta_j'\mid C]\ge\E\bigl[\widetilde C_1-\widetilde C_j\bigm| C\bigr]
 \ge\bigl(1+\gamma_0\,\alpha(C_1)\bigr)\Delta_j\ge\bigl(1+\gamma_0\,\alpha(C_1)\bigr)\Delta .
\]
If \(C_j=0\), then \(C_j'=0\) and \(\Delta_j'=C_1'\), so (c) and \(\Delta\le C_1\) give
\(\E[\Delta_j'\mid C]=\rho C_1\ge C_1+\gamma_0\,\alpha(C_1)\,\Delta\ge(1+\gamma_0\,\alpha(C_1))\Delta\).

(e) Let \(C_1\ge3n/4\), so that \(p_1=\max_\ell p_\ell\). The rule that outputs the most frequent of its
first three observations, breaking ties uniformly at random, reads \(h\ge3\) labels and is
label-equivariant; let \(r\) be its output law on \(p\). If at most one of the first three observations
differs from \(1\), then \(1\) is the unique most frequent of them. Each observation differs from \(1\)
with probability \(U/n\), independently, so \(1-r_1\le3(U/n)^2\). By \cref{lem:compare}, \(q_1\ge r_1\),
and hence \(\E[U'\mid C]=n(1-q_1)\le n(1-r_1)\le3U^2/n\).
\end{proof}

\section{From one round to consensus}\label{sec:rounds}

Throughout this section, \(h\ge3\), \(C^{(t)}\) is the configuration of \hmaj in \pullh after \(t\)
rounds, \(p^{(t)}=C^{(t)}/n\) is the vector of opinion densities, and \(\mathcal F_t\) is the history up
to round \(t\). We write \(C^{(t)}_1\), \(\Delta^{(t)}\) and \(\Delta^{(t)}_j\) for the quantities of
\cref{sec:model} at round \(t\), and \(\rho_t=\rho(C^{(t)})\) when opinion \(1\) is a largest opinion of
\(C^{(t)}\). Labels are not reordered over time: opinion \(1\) is the initially largest opinion
throughout. By the transition law of \cref{sec:model}, given \(\mathcal F_t\), the configuration
\(C^{(t+1)}\) has the multinomial distribution with parameters \(n\) and \(q(p^{(t)})\), which depends on
the history only through \(C^{(t)}\). Hence \((C^{(t)})_{t\ge0}\) is a time-homogeneous Markov chain, and
each lemma below, stated for an arbitrary initial configuration, also applies from a stopping time,
conditionally on the history up to it (strong Markov property).

While fewer than \(3n/4\) agents hold opinion \(1\), every round multiplies the bias by
\(1+\Omega(\alpha(M_0))\) with high probability (\cref{lem:round}); then the remaining agents adopt
opinion \(1\) within \(O(\log n)\) rounds (\cref{lem:endgame}). This round-by-round argument follows the
one for \dejavuh in~\cite[Section~6.3]{damore2026dejavuminimalisticmechanismdistributed}, with
\cref{lem:one-round} in place of the one-round estimates for \dejavuh. With this argument, our bias condition is stronger
than that of \dejavuh~\cite[Theorem~1]{damore2026dejavuminimalisticmechanismdistributed} because of
the fluctuations: in \dejavuh only the agents that see a repetition can change opinion, and the variance of
one round carries the probability of this event as a
factor~\cite[Lemma~18]{damore2026dejavuminimalisticmechanismdistributed}, whereas in \hmaj every agent
resamples, and the variance can be of order \(C_1\) (\cref{lem:one-round}(b) bounds it by \(\rho C_1\)). In this section,
\(\gamma_0\) is the absolute constant of \cref{lem:one-round}.

\subsection{One round under the bias condition}

\begin{lemma}[One round under the bias condition]\label{lem:round}
For every \(c>0\) there is a constant \(\lambda_c\ge1\), depending only on \(c\), such that the following
holds for every \(\lambda\ge\lambda_c\). Let \(C\) be a configuration in which opinion \(1\) is the unique
largest opinion, with \(C_1<3n/4\) and
\begin{equation}\label{eq:round-condition}
 \alpha(C_1)\,\Delta\ge\lambda\sqrt{C_1\log n}.
\end{equation}
Let \(C'\) be the configuration after one round of \hmaj in \pullh from \(C\), and let
\(\gamma=\gamma_0\,\alpha(C_1)\). With probability at least \(1-2n^{-(c+2)}\),
\[
 C_1'\ge C_1,\qquad\Delta'\ge\Bigl(1+\frac34\gamma\Bigr)\Delta,\qquad
 \alpha(C_1')\,\Delta'\ge\lambda\sqrt{C_1'\log n}.
\]
In particular, opinion \(1\) is the unique largest opinion of \(C'\).
\end{lemma}

\begin{proof}
Let \(K_c=2\sqrt{c+3}+2(c+3)\); we prove the lemma with \(\lambda_c=8K_c/\gamma_0\).

\emph{Expected values.} Let \(\rho=\rho(C)\), and note that \(\gamma\le1\), as \(\gamma_0\le1\) and \(\alpha(C_1)\le1\). By \cref{lem:one-round}(d), for
every \(j\ne1\),
\[
 \E[\Delta_j'\mid C]\ge\max\{\rho,1+\gamma\}\,\Delta .
\]
By \cref{lem:one-round}(c),
\[
 \E[C_1'\mid C]=\rho C_1\ge C_1+\gamma\Delta .
\]

\emph{Fluctuations.} Since \(\Delta\le C_1\) and \(\alpha(C_1)\le1\), \eqref{eq:round-condition} gives
\(C_1\ge\lambda\sqrt{C_1\log n}\), that is, \(C_1\ge\lambda^2\log n\). Let \(y=(c+3)\log n\) and
\[
 D=\sqrt{4\rho C_1y}+2y .
\]
By \cref{lem:one-round}(b), each of \(C_1'\) and \(\Delta_j'\) (\(j\ne1\)) is, given \(C\), a sum of \(n\)
independent terms that differ from their means by at most \(2\), with variance at most \(2\rho C_1\).
\Cref{lem:bernstein} with \(J=2\) and \(V=2\rho C_1\) shows that each of these at most \(n\) variables
differs from its conditional mean by less than \(D\), except with probability at most
\(2e^{-y}=2n^{-(c+3)}\). By a union bound, all of them do so with probability at least
\(1-2n^{-(c+2)}\), and we work on this event from now on. Since \(\lambda\ge1\) and
\(C_1\ge\lambda^2\log n\), we have \(\log n\le\sqrt{C_1\log n}\); together with \(\rho\ge1\),
\eqref{eq:round-condition} and \(\lambda\ge8K_c/\gamma_0\), this gives
\[
 D\le K_c\sqrt\rho\,\sqrt{C_1\log n}\le K_c\sqrt\rho\,\frac{\alpha(C_1)\Delta}\lambda
 =\frac{K_c}{\gamma_0\lambda}\,\sqrt\rho\,\gamma\Delta\le\frac{\sqrt\rho\,\gamma\Delta}8 .
\]

\emph{Growth of the bias and of the count.} If \(\rho\le4\), then \(D\le\gamma\Delta/4\), and, using
\(\Delta\le C_1\),
\[
 \Delta'\ge\max\Bigl\{\rho-\frac\gamma4,1+\frac34\gamma\Bigr\}\Delta,\qquad
 C_1\le C_1+\gamma\Delta-\frac{\gamma\Delta}4\le C_1'\le\Bigl(\rho+\frac\gamma4\Bigr)C_1 .
\]
If \(\rho>4\), then \(D\le\rho\Delta/16\), because \(\gamma\le1\) and \(\sqrt\rho\le\rho/2\); hence
\[
 \Delta'\ge\frac{15}{16}\rho\Delta,\qquad C_1\le\frac{15}{16}\rho C_1\le C_1'\le\frac{17}{16}\rho C_1 .
\]
In both cases \(C_1'\ge C_1\) and \(\Delta'\ge(1+\frac34\gamma)\Delta\). In particular \(\Delta'>0\), so
opinion \(1\) is the unique largest opinion of \(C'\).

\emph{The condition~\eqref{eq:round-condition} is preserved.} We have
\(\alpha(x)/\sqrt x=\min\{h^2\sqrt x/n,1/\sqrt x\}\), and, when \(x\) increases to \(x'\), the first term
increases and the second decreases by the factor \(\sqrt{x/x'}\). Hence \(\alpha(x')/\sqrt{x'}\ge\sqrt{x/x'}\,\alpha(x)/\sqrt x\)
for \(x'\ge x>0\). As \(C_1'\ge C_1\), this and~\eqref{eq:round-condition} give
\[
 \frac{\alpha(C_1')\,\Delta'}{\sqrt{C_1'}}\ge\frac{\Delta'}\Delta\sqrt{\frac{C_1}{C_1'}}\cdot
 \frac{\alpha(C_1)\,\Delta}{\sqrt{C_1}}\ge\frac{\Delta'}\Delta\sqrt{\frac{C_1}{C_1'}}\cdot\lambda\sqrt{\log n},
\]
so it suffices to show that \(\Delta'/\Delta\ge\sqrt{C_1'/C_1}\). We distinguish three cases.
\begin{itemize}[leftmargin=*]
\item If \(\rho\le1+\gamma\), then \(C_1'/C_1\le1+\frac54\gamma\), and
\[
 \sqrt{\frac{C_1'}{C_1}}\le1+\frac58\gamma\le1+\frac34\gamma\le\frac{\Delta'}\Delta .
\]
\item If \(1+\gamma<\rho\le4\), let \(x=\rho-\gamma/4\), so that \(x>1+\frac34\gamma\) and
\(C_1'/C_1\le x+\gamma/2\). Since \(x^2-x=x(x-1)\ge\frac34\gamma\ge\frac\gamma2\),
\[
 \sqrt{\frac{C_1'}{C_1}}\le\sqrt{x+\frac\gamma2}\le x\le\frac{\Delta'}\Delta .
\]
\item If \(\rho>4\), then
\[
 \sqrt{\frac{C_1'}{C_1}}\le\sqrt{\frac{17}{16}\rho}\le\frac{15}{16}\rho\le\frac{\Delta'}\Delta,
\]
where the middle inequality holds because \(\rho\ge272/225\). \qedhere
\end{itemize}
\end{proof}

\subsection{The endgame}

\begin{lemma}[Endgame]\label{lem:endgame}
For every \(c>0\) there is \(n_c\), depending only on \(c\), such that the following holds for
\(n\ge n_c\). If \(C^{(0)}_1\ge3n/4\), then, with probability at least \(1-n^{-(c+1)}\), the
configuration is a consensus on opinion \(1\) in round \(t_c=\lceil(c+3)\log_{4/3}n\rceil=O(\log n)\).
\end{lemma}

\begin{proof}
Let \(U_t=n-C^{(t)}_1\), and let \(\mathcal A_t\) be the event that \(U_s\le n/4\) for all \(s\le t\)
(with \(\mathcal A_{-1}\) the sure event). On \(\mathcal A_t\), \cref{lem:one-round}(e) gives
\(\E[U_{t+1}\mid\mathcal F_t]\le3U_t^2/n\le\frac34U_t\). Since \(\mathcal A_t\subseteq\mathcal A_{t-1}\)
and \(U_t\ge0\),
\[
 \E\bigl[U_{t+1}\one_{\mathcal A_t}\bigr]\le\frac34\E\bigl[U_t\one_{\mathcal A_t}\bigr]
 \le\frac34\E\bigl[U_t\one_{\mathcal A_{t-1}}\bigr],
\]
and by induction \(\E[U_t\one_{\mathcal A_{t-1}}]\le(3/4)^tU_0\le n(3/4)^t\). Moreover, on \(\mathcal A_t\),
given \(\mathcal F_t\), \(U_{t+1}\) is a sum of \(n\) independent indicators with mean at most
\(3U_t^2/n\le3n/16\) and variance at most \(3n/16\); \cref{lem:bernstein} with \(J=1\), \(V=3n/16\) and
\(y=n/256\) gives \(\sqrt{2Vy}+Jy<n/16\), so \(\Pr(U_{t+1}>n/4\mid\mathcal F_t)\le2e^{-n/256}\) on
\(\mathcal A_t\). Hence
\[
 \Pr(U_{t_c}\ge1)\le\Pr(\mathcal A_{t_c-1}^c)+\E\bigl[U_{t_c}\one_{\mathcal A_{t_c-1}}\bigr]
 \le2t_ce^{-n/256}+n\Bigl(\frac34\Bigr)^{t_c}\le2t_ce^{-n/256}+n^{-(c+2)}\le n^{-(c+1)}
\]
for \(n\ge n_c\).
\end{proof}

\subsection{Proof of \texorpdfstring{\cref{thm:main}}{Theorem 1}}

Fix \(c>0\), let \(\lambda=\lambda_c\) be as in \cref{lem:round}, and let
\[
 T_1=\biggl\lceil\frac{\log n}{\log(1+3\gamma_0\,\alpha(M_0)/4)}\biggr\rceil .
\]
Since \(\log(1+x)\ge x/2\) for \(0\le x\le1\), and \(1/\alpha(M_0)=\max\{1,n/(h^2M_0)\}\),
\[
 T_1\le1+\frac8{3\gamma_0\,\alpha(M_0)}\log n=O\Bigl(\Bigl(1+\frac n{h^2M_0}\Bigr)\log n\Bigr).
\]
Moreover, \(\alpha(M_0)\ge\min\{9/n,1\}\ge1/n\), as \(h\ge3\) and \(M_0\ge1\), so
\(T_1\le1+3n\log n/\gamma_0\). Let \(n_c'\ge3\) be at least the \(n_c\) of \cref{lem:endgame} and such
that \(2(1+3n\log n/\gamma_0)\,n^{-(c+2)}\le n^{-c}/2\) for all \(n\ge n_c'\). We prove the theorem with
\(L_c=\max\{\lambda,L^*_c\}\), where \(L^*_c\) is such that \((L^*_c)^2>n/\log n\) for all \(2\le n<n_c'\).
As noted after \cref{thm:main}, the bias condition reads \(\alpha(M_0)\Delta_0\ge L_c\sqrt{M_0\log n}\)
and implies \(n\ge M_0\ge L_c^2\log n\); hence \(n\ge n_c'\).

Let \(\tau\) be the first round with \(C^{(\tau)}_1\ge3n/4\) (so \(\tau=0\) if \(M_0\ge3n/4\)). Let
\(\mathcal B\) be the event that, for some round \(t<\min\{\tau,T_1\}\), the configuration \(C^{(t)}\)
satisfies the hypotheses of \cref{lem:round} but its conclusions fail for the transition from
\(C^{(t)}\) to \(C^{(t+1)}\). Applying \cref{lem:round} conditionally on \(\mathcal F_t\) for each
\(t<T_1\), and a union bound, gives \(\Pr(\mathcal B)\le2T_1n^{-(c+2)}\le n^{-c}/2\), by the choice of
\(n_c'\).

On the complement of \(\mathcal B\), we show by induction that, for \(0\le t\le\min\{\tau,T_1\}\), opinion
\(1\) is the unique largest opinion of \(C^{(t)}\) and
\begin{equation}\label{eq:induction}
 C^{(t)}_1\ge M_0,\qquad\alpha\bigl(C^{(t)}_1\bigr)\,\Delta^{(t)}\ge\lambda\sqrt{C^{(t)}_1\log n},\qquad
 \Delta^{(t)}\ge\Bigl(1+\frac34\gamma_0\,\alpha(M_0)\Bigr)^t\Delta_0 .
\end{equation}
For \(t=0\), this is the bias condition, as \(\lambda\le L_c\). Let \(t<\min\{\tau,T_1\}\) and
assume~\eqref{eq:induction} at round \(t\). Since \(t<\tau\), we have \(C^{(t)}_1<3n/4\), so \(C^{(t)}\)
satisfies the hypotheses of \cref{lem:round}, and its conclusions hold outside \(\mathcal B\). They
give~\eqref{eq:induction} at round \(t+1\), because \(\alpha\) is nondecreasing and \(C^{(t)}_1\ge M_0\),
so that the factor of \cref{lem:round} is
\(1+\frac34\gamma_0\,\alpha(C^{(t)}_1)\ge1+\frac34\gamma_0\,\alpha(M_0)\).

Outside \(\mathcal B\), we have \(\tau\le T_1\): otherwise~\eqref{eq:induction} at \(t=T_1\) and
\(\Delta_0\ge1\) would give \(\Delta^{(T_1)}\ge n\), whereas \(\Delta^{(T_1)}\le C^{(T_1)}_1<3n/4\).

At round \(\tau\), a stopping time, at least \(3n/4\) agents hold opinion \(1\). By the strong Markov
property and \cref{lem:endgame}, the configuration is a consensus on opinion \(1\) \(t_c\) rounds later,
except with probability at most \(n^{-(c+1)}\). Hence, except with probability at most
\(n^{-c}/2+n^{-(c+1)}\le n^{-c}\), consensus on opinion \(1\) is reached within \(T_1+t_c\) rounds. As
\(n\ge3\), the bounds on \(T_1\) and \(t_c\) give \(T_1+t_c\le A_c(1+n/(h^2M_0))\log n\) for a constant
\(A_c\) depending only on \(c\). Consensus is absorbing, which completes the proof of \cref{thm:main}.
\qed

\section*{AI usage statement}

Large language models from OpenAI, Anthropic and Google were used throughout this work, both to
develop the proofs and to draft the text. The authors are fully responsible
for all content of the note.

{\small
\bibliographystyle{alpha}
\bibliography{\hmajdir refs}

\newcommand{\etalchar}[1]{$^{#1}$}
\begin{thebibliography}{dDG{\etalchar{+}}26b}

\bibitem[BCE{\etalchar{+}}17]{BerenbrinkCEKMN17}
Petra Berenbrink, Andrea Clementi, Robert Els{\"{a}}sser, Peter Kling, Frederik
  Mallmann{-}Trenn, and Emanuele Natale.
\newblock Ignore or comply?: On breaking symmetry in consensus.
\newblock In {\em {ACM} Symposium on Principles of Distributed Computing,
  {PODC}}, pages 335--344, 2017.

\bibitem[BCN{\etalchar{+}}16]{BecchettiCNPT16}
Luca Becchetti, Andrea Clementi, Emanuele Natale, Francesco Pasquale, and Luca
  Trevisan.
\newblock Stabilizing consensus with many opinions.
\newblock In {\em Twenty-Seventh Annual {ACM-SIAM} Symposium on Discrete
  Algorithms, {SODA}}, pages 620--635, 2016.

\bibitem[BCN{\etalchar{+}}17]{becchettiSimpleDynamicsPlurality2017}
Luca Becchetti, Andrea Clementi, Emanuele Natale, Francesco Pasquale, Riccardo
  Silvestri, and Luca Trevisan.
\newblock Simple dynamics for plurality consensus.
\newblock {\em Distributed Computing}, 30(4):293--306, 2017.

\bibitem[BCN20]{becchettiConsensusDynamicsOverview2020}
Luca Becchetti, Andrea Clementi, and Emanuele Natale.
\newblock Consensus dynamics: An overview.
\newblock {\em ACM SIGACT News}, 51(1):58--104, 2020.

\bibitem[BEM59]{bechhofer1959single}
Robert~E. Bechhofer, Salah~E. Elmaghraby, and Norman Morse.
\newblock A single-sample multiple-decision procedure for selecting the
  multinomial event which has the highest probability.
\newblock {\em Ann. Math. Statist.}, 30(1):102--119, 1959.

\bibitem[BG52]{bahadur1952impartial}
Raghu~Raj Bahadur and Leo~A. Goodman.
\newblock Impartial decision rules and sufficient statistics.
\newblock {\em Ann. Math. Statist.}, 23(4):553--562, 1952.

\bibitem[BLM13]{boucheron2013concentration}
St{\'e}phane Boucheron, G{\'a}bor Lugosi, and Pascal Massart.
\newblock {\em Concentration Inequalities: A Nonasymptotic Theory of
  Independence}.
\newblock Oxford University Press, 2013.

\bibitem[CER14]{cooper2014power}
Colin Cooper, Robert Els{\"{a}}sser, and Tomasz Radzik.
\newblock The power of two choices in distributed voting.
\newblock In {\em Automata, Languages, and Programming - 41st International
  Colloquium, ICALP 2014}, pages 435--446, 2014.

\bibitem[CMR{\etalchar{+}}25]{CooperMRSS25}
Colin Cooper, Frederik Mallmann{-}Trenn, Tomasz Radzik, Nobutaka Shimizu, and
  Takeharu Shiraga.
\newblock Asynchronous 3-{M}ajority dynamics with many opinions.
\newblock In {\em 2025 Annual {ACM-SIAM} Symposium on Discrete Algorithms,
  {SODA}}, pages 4095--4131, 2025.

\bibitem[D'A26]{darchivio2026consensus}
Niccol{\`{o}} D'Archivio.
\newblock Consensus time in 3-{M}ajority and 2-{C}hoices is determined by the
  maximum initial opinion density.
\newblock In {\em 40th International Symposium on Distributed Computing,
  {DISC}}, 2026.
\newblock To appear. arXiv:2606.11778.

\bibitem[dDG{\etalchar{+}}26a]{damore2026dejavuba}
Francesco d'Amore, Niccol{\`{o}} D'Archivio, George Giakkoupis,
  Fr{\'{e}}d{\'{e}}ric Giroire, and Emanuele Natale.
\newblock Brief announcement: {D{\'{e}}j{\`{a}}Vu}: A minimalistic mechanism
  for distributed plurality consensus.
\newblock In {\em Proceedings of the ACM Symposium on Principles of Distributed
  Computing, PODC 2026, Egham, United Kingdom, July 6-10, 2026}, pages
  437--440, 2026.

\bibitem[dDG{\etalchar{+}}26b]{damore2026dejavuminimalisticmechanismdistributed}
Francesco d'Amore, Niccolò D'Archivio, George Giakkoupis, Frédéric Giroire,
  and Emanuele Natale.
\newblock {D{\'e}j{\`a}Vu}: A minimalistic mechanism for distributed plurality
  consensus.
\newblock arXiv:2604.03648v3, 2026.

\bibitem[dDGN25]{damoreHMajorityDynamicsMany2025}
Francesco d'Amore, Niccol{\`{o}} D'Archivio, George Giakkoupis, and Emanuele
  Natale.
\newblock On the $h$-{M}ajority dynamics with many opinions.
\newblock In {\em 39th International Symposium on Distributed Computing,
  {DISC}}, pages 27:1--27:24, 2025.

\bibitem[DGGN26]{darchivio2026dejavufirst}
Niccol{\`o} D'Archivio, George Giakkoupis, Fr{\'e}d{\'e}ric Giroire, and
  Emanuele Natale.
\newblock {D{\'e}j{\`a}Vu}: Plurality consensus by first repetition.
\newblock Manuscript, 2026.

\bibitem[DGM{\etalchar{+}}11]{DoerrGMSS11}
Benjamin Doerr, Leslie~Ann Goldberg, Lorenz Minder, Thomas Sauerwald, and
  Christian Scheideler.
\newblock Stabilizing consensus with the power of two choices.
\newblock In {\em Proceedings of the 23rd Annual {ACM} Symposium on Parallelism
  in Algorithms and Architectures, {SPAA}}, pages 149--158. {ACM}, 2011.

\bibitem[dRS{\etalchar{+}}26]{damore2026nearlyoptimal}
Francesco d'Amore, Joel Rybicki, Nobutaka Shimizu, Takeharu Shiraga, and Robin
  Vacus.
\newblock Nearly-optimal time-sample trade-off in consensus dynamics.
\newblock arXiv:2610.40562v2, 2026.

\bibitem[EFK{\etalchar{+}}16]{elsasser2016efficient}
Robert Els{\"{a}}sser, Tom Friedetzky, Dominik Kaaser, Frederik Mallmann-Trenn,
  and Horst Trinker.
\newblock Efficient $k$-party voting with two choices.
\newblock arXiv:1602.04667, 2016.

\bibitem[GL18]{GhaffariL18}
Mohsen Ghaffari and Johannes Lengler.
\newblock Nearly-tight analysis for 2-{C}hoice and 3-{M}ajority consensus
  dynamics.
\newblock In {\em 2018 {ACM} Symposium on Principles of Distributed Computing,
  {PODC}}, pages 305--313, 2018.

\bibitem[HP01]{hassin2001distributed}
Yehuda Hassin and David Peleg.
\newblock Distributed probabilistic polling and applications to proportionate
  agreement.
\newblock {\em Inf. Comput.}, 171(2):248--268, 2001.

\bibitem[Leh66]{lehmann1966theorem}
Erich~L. Lehmann.
\newblock On a theorem of {B}ahadur and {G}oodman.
\newblock {\em Ann. Math. Statist.}, 37(1):1--6, 1966.

\bibitem[SS25]{Shimizu2025}
Nobutaka Shimizu and Takeharu Shiraga.
\newblock 3-{M}ajority and 2-{C}hoices with many opinions.
\newblock In {\em {{ACM Symposium}} on {{Principles}} of {{Distributed
  Computing}}, PODC}, pages 207--217, 2025.

\end{thebibliography}
}

\end{document}